\documentclass[preprint2]{aastex}

\shorttitle{Core depletion}
\shortauthors{Alister W.\ Graham}

\begin{document}

\title{Core depletion from coalescing supermassive black holes}

\author{Alister W.\ Graham}
\affil{Department of Astronomy, University of Florida, Gainesville, FL, USA}
\email{Graham@astro.ufl.edu}

\begin{abstract}
New measurements of the stellar-mass deficits at the centers of
luminous elliptical galaxies are presented.  These were derived
considering the following observational facts. Firstly, ``core''
galaxies, which are thought to have had their inner region depleted
from the coalesence of supermassive black holes, show an abrupt
downward deviation of their inner light-profile relative to their
outer S\'ersic profile.  Second, ``power-law'' galaxies, having
undisturbed profiles and no partially depleted core, have inner
light-profiles that display no departure from the inward extrapolation
of their outer S\'ersic profile.  The central stellar deficits have
therefore been derived from the difference in flux between the {\sl
HST}-observed galaxy light-profiles and the inward extrapolation of
each galaxy's outer S\'ersic profile.  This approach gives flux
deficits $\sim$0.1\% 
of the total galaxy light, and mass deficits that
are $\sim$2 
times each galaxy's central supermassive black hole mass.  These
results are in agreement with the theoretical expectations of mass
ejection from binary black hole mergers and also with popular $\Lambda$CDM
models of hierarchical galaxy formation.  It is also explained why
this result is some 10 times smaller than current observational
estimates of the central mass deficit, and therefore implies a merger
history for giant elliptical galaxies that is one order of magnitude
less violent than previously suggested.
\end{abstract}


\section{Introduction}

The collisional construction of galaxies from the merger of lesser
galaxies is
thought to be a common occurrence in the Universe.  Coupled with the
presence of a supermassive black hole (SMBH) at the heart of most
galaxies (Kormendy \& Richstone 1995; Magorrian et al.\ 1998,
Richstone et al.\ 1998),
%
%
dissipationless mergers have been proposed to explain the damaged
nuclei in giant elliptical galaxies (e.g., Lauer et al.\ 1995, Faber
et al.\ 1997, Rest et al.\ 2001).  Although some galaxy
``core-depletion'' is due to the SMBH(s) dining on stars that venture
to close (e.g., Magorrian \& Tremaine 1999; Zhao, Haehnelt, \& Rees,
2002; Yu 2003),
it is primarily from the gravitational slingshot
effect that the coalescing SMBHs --- from the pre-merged galaxies ---
have on stars while they themselves sink to the bottom of the
potential well of the newly wed galaxy (Begelman, Blandford, \& Rees
1980; Ebisuzaki, Makino, \& Okumura 1991; Makino \& Ebisuzaki 1996;
Quinlan 1996; Quinlan \& Hernquist 1997).

Theory predicts that the orbital decay of two such SMBHs should eject
a core mass roughly equal to the combined black hole masses
(Ebisuzaki, Makino, \& Okumura 1991; Milosavljevi\'c \& Merritt 2001).
Current measurements of the central stellar deficit are an order of
magnitude larger than the central SMBH mass, suggesting that most
elliptical galaxies have undergone multiple ($\approx$8-10)
major-mergers (Milosavljevi\'c \& Merritt 2001; Milosavljevi\'c et
al.\ 2002; Ravindranath, Ho, \& Filippenko 2002).  This result,
however, is at odds with popular models of hierarchial structure
formation, which predict an average of only 1 (dissipationless)
major-merger event for luminous elliptical galaxies (Haehnelt \&
Kauffmann 2002; Volonteri, Madau, \& Haardt 2003).

Recent advances in our understanding of galaxy structure have provided
a new framework in which to think about, and measure, central mass
deficits.  It is now known that the so-called ``power-law'' galaxies
--- understood not to have partially depleted cores, nor experienced a
(major) dissipationless merger event --- have an undisturbed S\'ersic
$R^{1/n}$ profile over their {\it entire} radial extent.  That is,
their inner light-profiles show no deviation relative to their outer
$R^{1/n}$ profile (Trujillo et al. 2004).  On the other hand, the more
luminous ($M_B<-20.5$ mag) ``core'' galaxies display a clear
flattening of their inner light-profile relative to the inward
extrapolation of their outer S\'ersic profile (Graham et al.\ 
2003b, 2004; Graham \& Guzm\'an 2003; Trujillo et al.\ 2004).  The
`break' in the profiles where this transition occurs marking the 
boundary of their relatively unpopulated cores. 

With this new knowledge of what galaxy profiles look like, this paper
considers the form of the core-less galaxy light-profiles when
representing the {\it original} stellar distribution of the disturbed
profiles, in order to compute the central stellar deficit and hence
the level of damage to a galaxy's core.
%

\section{Method \& Results} 
\subsection{Central stellar deficits}

We proceed by quantifying the central stellar deficit as the flux
deficit relative to the inward extrapolation of the smoothly curving,
stellar distribution outside of any possible, partially depleted core.
This approach, therefore, does not assign {\it any} mass deficit to
``power-law'' galaxies, whose inner light-profiles display no clear
downward deviation from their outer S\'ersic light-profiles.  Such a
qualitative description can be placed on a quantitative footing
through employment of the ``core-S\'ersic'' light-profile model
(Graham et al.\ 2003b), applied in Figure 1 to the 
``core'' galaxy NGC~3348.  This model consists of an inner power-law
and an outer S\'ersic function.  In practice (Trujillo et al.\ 2004),
the transition at the ``break radius'' is sharp, providing a
5-parameter function\footnote{The functional form of the 
complete core-S\'ersic model can be seen in Graham et al.\ 2003b 
and Trujillo et al.\ 2004.}
capable of describing the entire radial extent of galaxies with cores.
%
%

The central flux deficit 
is obtained by differencing the luminosity $L(r)=\int_0^r2\pi
rI(r) dr$ within $r=r_b$ of a) the inwardly-extrapolated outer
S\'ersic profile 
\begin{equation}
L_{\rm Ser}(r=r_b)=I_e r_e^2 2\pi n\frac{e^{b_n}}{(b_n)^{2n}}\Gamma_{\rm I} (2n,b_n(r_b/r_e)^{1/n}),
\end{equation}
and b) a power-law light-profile with slope $\gamma$ matching the 
observed inner profile slope, and intercepting the S\'ersic model 
at $r=r_b$, 
\begin{equation}
L_{\rm p-law}(r=r_b) = 2\pi r_b^2I_b/(2-\gamma). 
\end{equation}
The term $I_{e}$ is the intensity of the S\'ersic profile at the
effective radius $r_e$ enclosing half the galaxy light.  The exponent
`$1/n$' describes the curvature of the light-profile, and $b_n$ is 
simply a function of $n$ such that $\Gamma(2n)=2\Gamma_{\rm I}(2n,b_n)$, 
%
where $\Gamma$ and 
$\Gamma_{\rm I}$ are the complete and incomplete gamma functions 
respectively.  The intensity at $r_b$ is denoted by $I_b$. 
%

This procedure has been applied here to the 7 bona fide core galaxies
from Trujillo et al.\ (2004)\footnote{Clarifying note: The structural
parameters tabulated in Trujillo et al.\ (2004) were obtained using the
approximation $b_n \approx 1.999n-0.3271$, and the 
$r_e$ values are for each galaxy's outer $R^{1/n}$ profile as if it had
no partially depleted core.}.  The resultant flux deficits are
0.07--0.7\% of the total galaxy flux.
Allowing for the fact that these galaxies are at a range of distances from
us, the {\it apparent} magnitude differences 
$m(r=r_b)=-2.5\log [L_{\rm Ser}-L_{\rm p-law}]$  
were converted into {\it absolute} magnitudes ($M$) 
using the galaxy distance estimates from Tonry et al.\ (2001)
and then converted into units of solar flux using an absolute $R$-band
magnitude for our Sun of $M_{\bigodot}$=4.46 mag (Cox 2000).  
Finally, these values were transformed into solar masses assuming a 
stellar mass-to-light ratio of 3.0 (Worthey 1994). 
Such a ratio is representative of an evolved (i.e.,
faded) 12 billion year old, single stellar population 
observed through an $R$-band optical filter. 
The mass deficits $M_{\rm def}$ are given in Table 1. 
%
%
%
%
%

\subsection{Constraints on the number of dry mergers}

Models of hierarchical structure formation predict that galaxies will
collide; indeed, this phenomenon has been observed for many years.
%
%
Theoretical expectations for the ejected core mass, after the violent
unification of galaxies containing SMBHs, scale as 0.5-2$NM_{\rm bh}$,
where $M_{\rm bh}$ is the {\it final} BH mass and $N$ is the number of
merger events (Milosavljevi\'c et al.\ 2002).  The variable term at
the front is because equal-mass mergers scour out more stars than a
collision involving a lesser-mass, secondary galaxy.  Thus, by knowing
a galaxy's central stellar deficit (Table 1), and its black hole mass
$M_{\rm bh}$, one can place constraints on the extent of this merger
process.

We have estimated the SMBH mass of each galaxy using two techniques.
First, we employed the $M_{\rm bh}$--$\sigma$ relation of Gebhardt et
al.\ (2000), in which $M_{\rm bh}$ is derived from the galaxy 
velocity dispersion $\sigma$.  Using the steeper $M_{\rm
bh}$--$\sigma$ relation of Ferrarese \& Merritt (2000) had little 
difference because of the different zero-points in these two relations. 
The second SMBH mass estimate came from the independent $M_{\rm bh}$--$n$
relation (Graham et al.\ 2001, 2003a).  This relation is as strong as the $M_{\rm
bh}$--$\sigma$ relation and has the same small degree of scatter, but
has the advantage that it is a purely photometric technique, requiring
only (uncalibrated) images rather than (telescope-time expensive) galaxy
spectra.
Figure 2 shows each galaxy's depleted core mass plotted against both
estimates of the SMBH mass.  The ratio of the stellar mass deficit to
central black hole mass has a mean $\pm$ average deviation of
$2.4\pm0.7$ (panel a, $M_{\rm bh}$--$n$) and $2.1\pm1.1$ (panel b,
$M_{\rm bh}$--$\sigma$), consistent with these galaxies having
experienced one major (i.e., equal-mass) dry merger event.


\section{Discussion}

Support for the above mass deficits comes from the agreement with the
merger simulations of Makino \& Ebisuzaki (1996), the 
(cusp regeneration) hierarchial merger models of Volonteri et al.\ 
(2003), and the theoretical expectations of Haehnelt \&
Kauffmann (2002).  Although the latter Authors predicted a median
number of equal mass mergers for faint (power-law) and bright (core)
elliptical galaxies of 1 and 3 respectively, the number of major
mergers since the last collision that involved substantial gas
accretion ($M_{\rm gas}>M_{\rm bh}$) is 0 and 1 respectively.  The
presence of gas is important because it dilutes the wrecking ball
action of the SMBHs on the stars because it, rather than the stars,
fosters the coalesence of the black holes and it can lead
to the creation of new stars (see, e.g., Zhao et al.\ 2002, and
references therein).  Therefore, our conclusions that power-law
galaxies do not have partially depleted cores from galaxy collisions,
and that the number of (dissipationless) major mergers 
producing
luminous galaxies, with $M_R\sim -22.5$ mag, is equal to about 1, are
supported by current cold dark matter models of hierarchial structure
formation.

Given $M_{\rm def} \sim M_{\rm bh}$, one may wonder if some 
depleted cores might have formed from the runaway merging of stars 
(e.g., Begelman \& Rees 1978; Quinlan \& Shapiro 1989)
within what may once have been the dense cores of massive ellipticals 
(Graham \& Guzm\'an 2003), 
%
%
rather than from the scattering of stars from coalescing black holes.
%
%
The first objection to such a process would be those cases in which
$M_{\rm def}$ is actually greater than $M_{\rm bh}$.  Such a mechanism
would also require a certain level of refinement, such as
re-populating the loss-cone, in order to explain the absence of
(resolved) cores in less luminous elliptical galaxies.  
The expected break radii --- derived from a centrally
depleted S\'ersic model --- are not observed amongst the ``power-law''
galaxies.  For example, if the 12 ``power-law'' galaxies in Trujillo
et al.\ (2004) had cores with inner power-law slopes ranging from
0.0--0.3, then, assuming $M_{\rm def}=M_{\rm bh}$, they should have
break radii of 0.17--0.5 arcseconds, which they do not.
%

It is pertinent to inquire why previous estimates of the central mass
deficit are larger than the values obtained here.  One reason is
that it had been assumed that every galaxy once had a steep
isothermal $\rho(r)\sim r^{-2}$ core before any merging black holes
wreaked their havoc.  There is, however, no observational evidence
that any such universal density-profile $\rho(r)$ exists, or once
existed amongst the ``power-law'' galaxies.  
In actuality, a luminosity-dependent range of inner profiles
shapes is now known to exist (Gebhardt et al.\ 1996; Graham \& Guzm\'an
2003; Balcells, Graham, \& Peletier 2004).  No single power-law
slope can be used to approximate the initial, undisturbed, stellar
distribution of all elliptical galaxies. 



Milosavljevi\^c et al.\ (2002), for example, had defined the onset
of partially depleted cores as the radius where the negative,
logarithmic gradient of the deprojected light-profiles (i.e., the
spatial, luminosity-density profiles) equaled 2.  That is, where the
observed slope matched that of the isothermal model.  They then
derived central, stellar mass deficits from the difference between the
observed density-profile and the inward extrapolation of the
isothermal model from this point.
This approach to estimating the depleted core mass is illustrated in
Figure 3, where we show both the observed light-profile and the
(deprojected) density-profile of the core-less, ``power-law'' galaxy NGC
5831.
Such a 
prescription encounters a number of difficulties. 
An extraordinary level of fine-tuning would be required to deplete
stars over the full radial extent of the initial isothermal core,
assuming one existed, but not beyond the final core radius, which
one assumes still has its original slope of -2 today.
Moreover, with such an approach, many of the ``power-law'' galaxies which
are not (traditionally) recognized as having a partially depleted core
will be assigned one.  Such theoretical core radii (and mass deficits)
are not only questionable, but excessively large --- sometimes
greater than 1 kpc --- and don't match the observed break radii in
core galaxies, which are invariably less than a few hundred parsecs
(see Table 1). 
%

A second approach to estimating the central deficit had been to use
break radii derived from the Nuker model (Lauer et al.\ 1995), and to
assume an isothermal core once existed inside of this radius.  With
such an approach the estimated central deficits are again too high
because of a) the excessively steep isothermal model that is assumed,
and b) because, as described in Graham et al.\ (2003b) and shown in
Trujillo et al.\ (2004), Nuker-derived break radii typically
overestimate the actual break radii by factors of 2 to 5. 

The mass deficits obtained with the above two methods are 3--30 times
larger than our values, and are shown in Figure 2 for comparison.  Our 
new measurements of the mass deficit are significantly (99.4\%, from
both a K-S test and Students' T-test) different to those values
obtained using the isothermal assumption, and reveal that the galactic
merger history of the Universe, at least for massive elliptical
galaxies, is roughly an order of magnitude less violent than previous
observational analyses (Milosavljevi\'c \& Merritt 2001; Milosavljevi\'c
et al.\ 2002; Ravindranath, Ho, \& Filippenko 2002) had suggested.


We plan on analyzing more galaxies, and measuring the ellipticity of
their evacuated core region.  
This may shed light on the orientation of the initial orbits of the
black holes, and allow one to explore any possible correlation with
the host galaxy ellipticity.  
A greater range of data will also allow
one to explore whether there is any trend between $M_{\rm def}/M_{\rm
bh}$ and galaxy magnitude.  Such a correlation may be expected if
bigger ``core'' galaxies have experienced more dissipationless merger
events than less luminous ``core'' galaxies.

\acknowledgements 
Support for proposal HST-AR-09927.01-A was provided
by NASA through a grant from the Space Telescope Science Institute,
which is operated by the Association of Universities for Research in
Astronomy, Inc., under NASA contract NAS5-26555.


\begin{deluxetable}{lcc|ccc|cc|cc}
\tablewidth{0pt}
\tablecaption{Galaxy data}
\tablehead{
\colhead{Galaxy} & \colhead{$\sigma$}    & \colhead{$M_{\rm gal}$} &  
\multicolumn{3}{c}{core-S\'ersic}  &  \multicolumn{2}{c}{Isothermal}  & \multicolumn{2}{c}{Nuker} 
\\[.2ex]
\colhead{}  & \colhead{(km s$^{-1}$)} &  \colhead{($R$-mag)}  & \colhead{$n$} &
\colhead{$r_b^{\prime\prime}$ (pc)} &  \colhead{$M_{\rm def}$}  &
\colhead{$r_b^{\prime\prime}$ (pc)} &  \colhead{$M_{\rm def}$}  &
\colhead{$r_b^{\prime\prime}$ (pc)} &  \colhead{$M_{\rm def}$} 
}
\startdata
NGC~2986 & 268 & -22.49 & 5.28 &  0.69 (97)  & 7.02 &    2.9 (410)  & 35.5 &    1.24 (174) & 26.7 \\
NGC~3348 & 237 & -22.76 & 3.81 &  0.35 (70)  & 3.01 &    2.5 (490)  & 66.1 &    0.99 (198) & 26.5 \\
NGC~4168 & 186 & -21.92 & 3.12 &  0.72 (108) & 1.21 &    8.3 (1250) & 38.0 &    2.02 (303) & 23.5 \\
NGC~4291 & 284 & -21.61 & 5.44 &  0.37 (47)  & 5.57 &    1.3 (170)  & 17.8 &    0.60 (076) & 16.7 \\
NGC~5557 & 253 & -23.08 & 4.37 &  0.23 (51)  & 2.13 &    2.0 (440)  & 41.7 &    1.21 (269) & 39.6 \\
NGC~5903 & 210 & -22.69 & 5.09 &  0.86 (141) & 8.44 &    4.7 (780)  & 38.0 &    1.59 (262) & 25.5 \\
NGC~5982 & 250 & -22.99 & 4.06 &  0.28 (57)  & 3.12 &    2.2 (450)  & 39.8 &    0.74 (151) & 23.5 \\
\enddata
\tablecomments{
Column 1: New General Catalog (NGC) numbers.
Column 2: Velocity dispersions $\sigma$ from Hypercat
(http://www-obs.univ-lyon1.fr/hypercat/).
Column 3: Absolute $R$-band
galaxy magnitudes $M_{\rm gal}$ derived from the best-fitting
(sharp transition) core-S\'ersic model; the parameters from which,
including the S\'ersic index $n$ and the break radii $r_b$ (in
arcseconds and parsecs), can be
found in Trujillo et al.\ 2004. 
The associated central mass deficits $M_{\rm def}$ (in
units of $10^8$ solar masses) have been derived from the difference in
flux between the observed light-profile and the inward extrapolation
of the outer S\'ersic profile.  The `Isothermal'
quantities have come directly from Milosavljevi\'c et al.\ 2002 who used the
technique illustrated in Figure 3b.  The `Nuker' break radii have come
from Rest et al.\ 2001, and the Nuker-derived mass deficits were obtained using 
equation 41 from Milosavljevi\'c \& Merritt 2001 with the required 
$\gamma$ values taken from Ravindranath et al.\ 2002. 
}
\end{deluxetable}


\begin{figure}
\epsscale{0.75}
\plotone{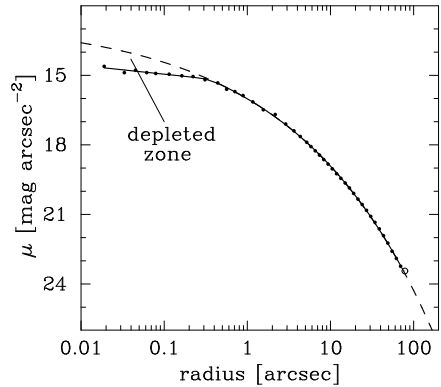}
\caption{
Observed, major-axis, $R$-band, surface brightness profile of
NGC~3348.  The solid line is the best-fitting core-S\'ersic model,
while the dashed line is the best-fitting S\'ersic model to the data
beyond the break radius $r_b$=0.35 arcseconds.  The flux deficit is
illustrated by the area designated as the `depleted zone',
corresponding to a mass deficit of 300 million solar masses.  Data
points from Trujillo et al.\ (2004), and supercede those shown 
in Graham et al.\ (2003b).
}
\label{fig1}
\end{figure}

\begin{figure}
\epsscale{0.75}
\plotone{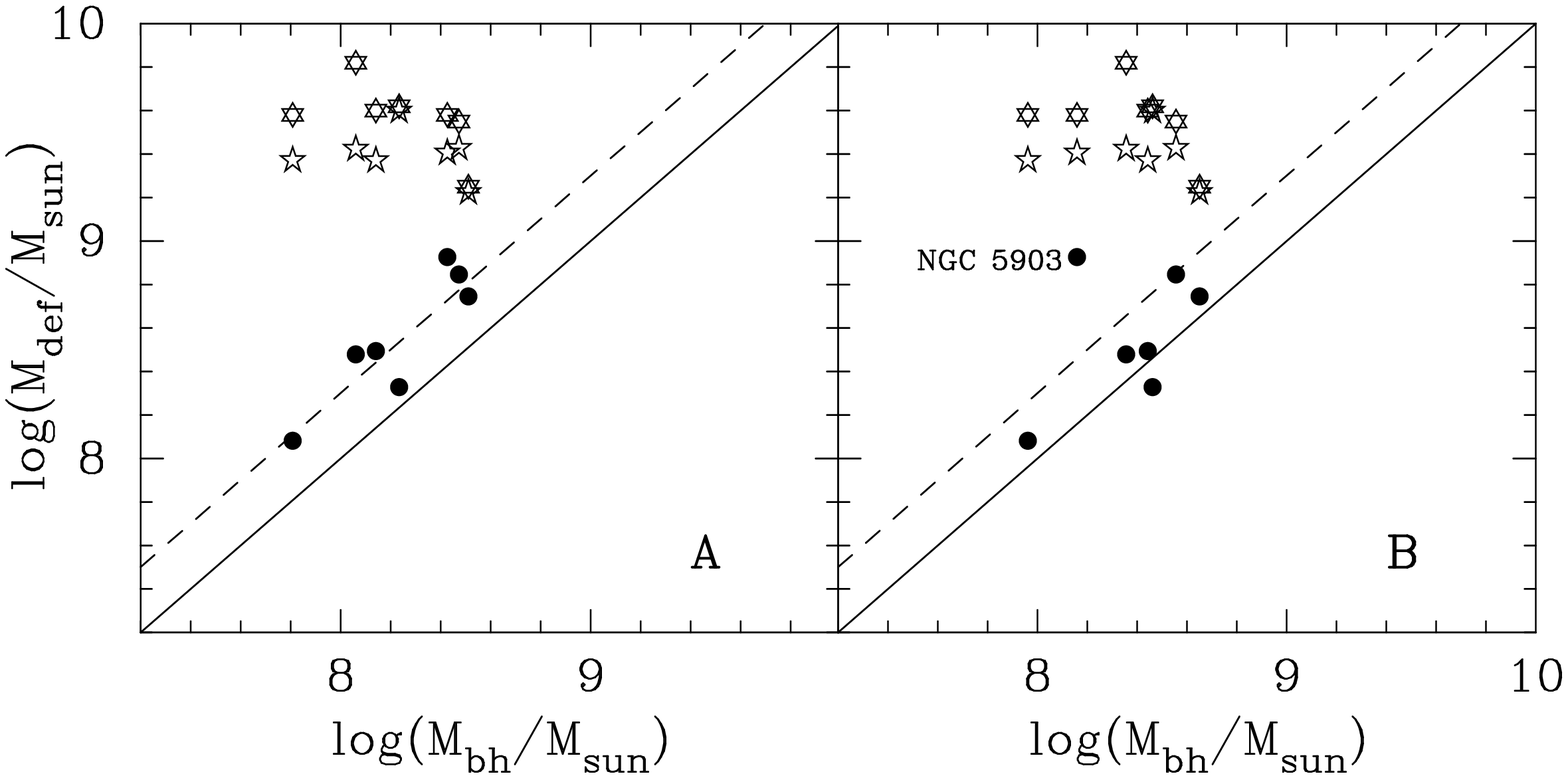}
\caption{
Central mass deficits evaluated from three techniques versus the
central black hole masses derived using a) the galaxy S\'ersic
index $n$ (Graham et al.\ 2001, 2003a), and b)
the velocity dispersion $\sigma$ (Gebhardt et al.\ 2000).
6--pointed stars: from Milosavljevi\'c et al.\ (2002) using the
method illustrated in Figure 3b.
5--pointed stars: derived using Nuker-model break-radii and equation 41 from 
Milosavljevi\'c \& Merritt (2001). 
Filled circles: derived here using the logic illustrated in Figure 1.
The solid line shows a one-to-one relation, the dashed line shows
$M_{\rm def}=2M_{\rm bh}$.
Typical errors for the points marked with a filled circle are roughly
$\log 2$ along both axes, stemming from a $\sim 20$\% and $\sim 15$\%
uncertainty on the value of $n$ and $\sigma$ respectively, and from
profile fitting and galaxy distance uncertainties.
}
\label{fig2}
\end{figure}

\begin{figure}
\epsscale{0.75}
\plottwo{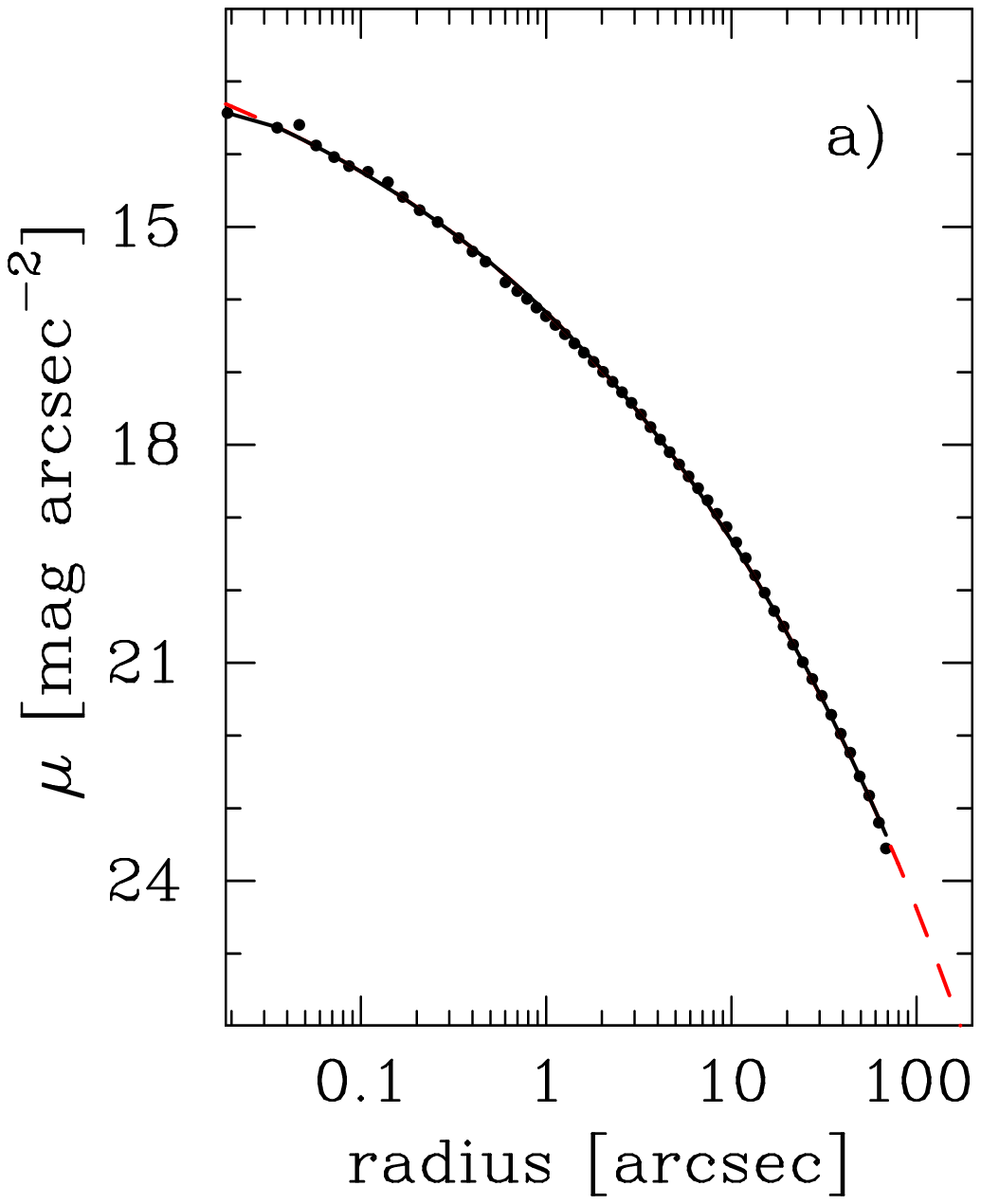}{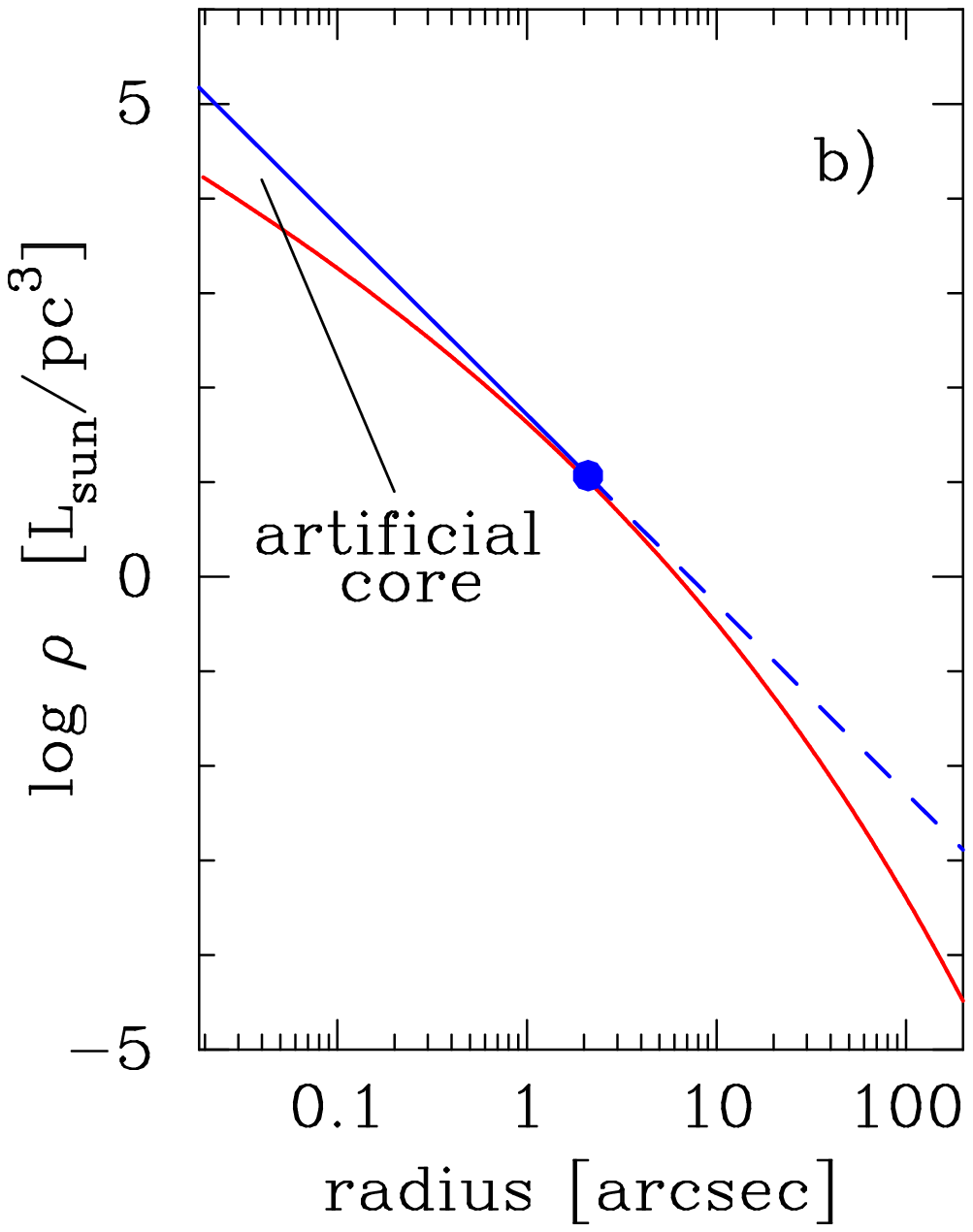}
\caption{a) 
Observed, major-axis, $R$-band, surface brightness profile
of the core-less galaxy NGC~5831 (data taken from Trujillo et al.\
2004).  A 3-parameter S\'ersic model (solid line) adequately describes
the stellar distribution on both the nuclear ($< 1^{\prime\prime}$)
and global scale; the dashed line shows the model extrapolated beyond
the data.  No characteristic downward break in the inner light-profile, that
would signify damage caused by merging SMBHs, is evident.
b) Spatial density profile (curved line) of NGC~5831
obtained by deprojecting the light-profile model in panel a).
Assuming an isothermal model, $\rho(r)\sim r^{-2}$, once existed
(straight line), one can assign a theoretical core radius $r_b$ ---
where the logarithmic density profile has a slope of -2 --- and a
central mass deficit.  Milosavljevi\'c et al.\ (2002) obtained
$r_b$=220 parsecs and $M_{\rm def}=5.6\times 10^8 M_{\bigodot}$.
}
\label{fig3}
\end{figure}

\end{document}